\documentclass[12pt]{utarticle}
\pdfoutput=1
\usepackage{authblk}
\usepackage{fullpage}
\usepackage[colorlinks=true,linkcolor=blue,urlcolor=blue,
  citecolor=blue]{hyperref}
\usepackage{slashed}
\usepackage{bbold}
\usepackage{amssymb,graphicx}
\usepackage{amsmath}
\usepackage[margin=0.5cm]{caption}
\usepackage{cite}
\usepackage{upgreek}
\usepackage[body={16.6cm, 22cm},right=1.5cm]{geometry}

\DeclareMathOperator{\arctanh}{arctanh}

\newcommand{\be}{\begin{equation}}
\newcommand{\ee}{\end{equation}}
\newcommand{\bea}{\begin{eqnarray}}
\newcommand{\eea}{\end{eqnarray}}
\newcommand{\bg}{\begin{gather}}
\newcommand{\eg}{\end{gather}}
\newcommand{\bseq}{\begin{subequations}}
\newcommand{\eseq}{\end{subequations}}

\def\e{{\rm e}}

\def\d{\partial}
\def\l{\left(}
\def\r{\right)}

\newcommand{\diagram}[1]{#1}
\begin{document}
\baselineskip=15.5pt

\begin{titlepage}
\begin{center}
{\Large\bf Looking for Integrability on the Worldsheet of Confining Strings}
 \\
\vspace{0.5cm}
{ \large
Patrick Cooper$^a$\footnote{\emailt{pjc370@nyu.edu}}, Sergei
Dubovsky$^{a,b}$\footnote{\emailt{sergei.dubovsky@gmail.com}},  Victor
Gorbenko$^a$\footnote{\emailt{gorbenko@nyu.edu}}, \\
\vspace{.25cm}
Ali Mohsen$^a$\footnote{\emailt{ahm302@nyu.edu}}, and Stefano
Storace$^a$\footnote{\emailt{ss5375@nyu.edu}}
}\\
\vspace{.45cm}
{\small  \textit{    $^a$Center for Cosmology and Particle Physics,\\ Department of Physics,
      New York University\\
      New York, NY, 10003, USA}}\\
      \vspace{.25cm}
      {\small  \textit{    $^b$Abdus Salam International Center for Theoretical Physics,\\ Strada
      Costiera 11, 34151, Trieste, Italy}}\\
\end{center}

\begin{abstract}
    We study  restrictions on scattering amplitudes  on the worldvolume of branes and strings (such
    as confining flux tubes in QCD) implied by the target space Poincar\'e symmetry. We focus on
    exploring the conditions for the string worldsheet theory to be integrable.  We prove that for a
    higher dimensional membrane the scattering amplitudes for the translational Goldstone modes
    (``branons") are double soft. At one-loop double softness is generically  violated  for the
    string worldsheet scattering as a consequence of collinear singularities. Violation of double
    softness implies in turn the breakdown of integrability. We prove that if branons are the only
    gapless degrees of freedom then the worldsheet integrability is compatible with target space
    Poincar\'e symmetry only if the number of space-time dimensions is equal to $D=26$ (a critical
    bosonic string), and for $D=3$. We extend the analysis to include massless worldsheet fermions,
    resulting from spontaneous breakdown of the target space supersymmetry.  We check that the
    tree-level integrability in this case is in one-to-one correspondence with the existence of
    a $\varkappa$-symmetric Green-Schwarz (GS) action. As a byproduct we show that at the leading
    order in the derivative expansion an $\mathcal{N}=1$ superstring without $\varkappa$-symmetry in
    $D=3,4,6,10$  dimensions exhibits an  accidental enhanced supersymmetry and is equivalent to  a
    $\varkappa$-symmetric $\mathcal{N}=2$ GS superstring.
\end{abstract}
\end{titlepage}
\section{Introduction}
String theory was born as a model of strong interactions \cite{Veneziano:1968yb}, and this is not a
coincidence.  The stringy nature of hadrons is convincingly proven both by linearly rising Regge
trajectories (see, e.g., \cite{Shifman:2007xn} for a recent compilation of meson spectra), and by
lattice simulations\footnote{See 
\url{http://www.physics.adelaide.edu.au/theory/staff/leinweber/VisualQCD/Nobel/} for animations.}.
Nevertheless the theory of the QCD confining string  (also known as the flux tube) still remains
elusive.

Remarkable progress in this direction was achieved recently for the $\mathcal{N}=4$ supersymmetric
cousin of QCD, where it turned out to be possible to find an exact integrable S-matrix on the flux
tube worldsheet, allowing one to solve the theory in the planar limit (for a review, and more
precise explanation of what ``solve" means here, see  \cite{Beisert:2010jr}; for most recent
developments and further later references see, e.g., \cite{Gromov:2014caa}).  Note, however, that
$\mathcal{N}=4$ supersymmetric Yang--Mills (SYM) is a conformal theory, and as such does not possess
confining flux tubes in a strict sense. Instead,  in this case it is more appropriate to think about
strings in the dual $AdS_5\times S^5$ space.

In addition, an approximate low energy integrability of the worldsheet theory proved to be a
powerful perturbative tool for calculating the spectra of confining strings in non-supersymmetric
gluodynamics \cite{Dubovsky:2013gi,Dubovsky:2014fma}. This new technique allowed for a solid
theoretical interpretation of the available high quality lattice data
\cite{Athenodorou:2010cs,Athenodorou:2013ioa}, and lead to the identification of the first massive
excited state on the worldsheet of the QCD string.

Of course, there is no reason to expect the worldsheet theory of the confining string in real
world QCD to be integrable. However, given the integrability of the $\mathcal{N}=4$ SYM string it is
tempting to look for genuinely confining theories with less supersymmetries  enjoying integrability
on the flux tube worldsheet, at least in the planar limit.

The goal of the present paper is to take a small step in this direction, by identifying necessary
conditions for integrability of  the flux tube theory. We take a bottom up approach and study the
constraints on the low energy worldsheet scattering imposed  by the Poincar\'e symmetry of the full
theory.

In section~\ref{sec:soft} we start with a few general remarks about the properties of worldvolume
scattering amplitudes for branes embedded into Minkowski space-time. From an effective field theory
point of view, a brane may be thought of as a system of Goldstone bosons---branons, corresponding to
the spontaneous breaking of translational invariance.  We prove that for a generic number of
worldvolume dimensions the branons' amplitudes are doubly soft---in the limit when one of the
scattering branons is soft, the amplitude vanishes as the second power of its energy. This is a natural
extension of the conventional soft pion theorems.

In section~\ref{sec:bosonic} we study whether worldsheet integrability is possible when branons  are
the only gapless worldsheet degrees of freedom. In other words, at low energies a flux tube is
described by a Nambu--Goto bosonic string in this case.  Then integrability is always present at the
tree-level.  The key observation behind our analysis is that string worldsheet amplitudes are
universal not only at tree-level, but also at one-loop \cite{Dubovsky:2012sh}. By direct
inspection of one-loop scattering we find that integrability is broken unless the string is
critical, $D=26$, or the dimensionality of the target space-time is $D=3$.  This breakdown of
integrability can be traced to the violation of the double softness property, as a consequence of
collinear singularities contributing to the corresponding Ward identities.  This proves the
following no-go theorem:

{\it Worldsheet scattering on the confining flux tube of a four-dimensional gauge theory can be
integrable only in the presence of additional gapless degrees of freedom.}

In section~\ref{sec:susy} we consider whether the no-go theorem may be avoided by introducing
additional massless degrees of freedom on the worldsheet. Here we restrict to the simplest
possibility, namely, we add goldstini---fermions which are massless as a consequence of non-linearly
realized target space supersymmetry.  We find that in this case even tree-level integrability is
not guaranteed any longer. If all supercharges are broken, tree level integrability holds only for
$\mathcal{N}=1$ supersymmetry in space-time dimensionalities $D=3,4,6,10$.  These are  the
dimensionalities where the $\varkappa$-symmetric Green--Schwarz (GS) superstring action can be
constructed \cite{Green:1983wt}. The GS superstring does exhibit unbroken supercharges and is integrable
at tree level both for $\mathcal{N}=1$ and  $\mathcal{N}=2$.

 This analysis makes slightly more precise the standard lore that classical superstrings exist only
 for the above special choices for the number of space-time dimensions and supercharges. This cannot
 be literally correct, given that at the effective field theory level one clearly can construct
 string actions corresponding to an arbitrary choice of supersymmetric coset. The accurate statement
 is that a {\it classically integrable} superstring action exists only for $\mathcal{N}=1,2$ and
 $D=3,4,6,10$. As for the bosonic string, the classical integrability survives at the quantum level
 only for the critical superstrings $D=10$, or at $D=3$.

 As a byproduct of this analysis we also prove the following. In principle, for $\mathcal{N}=1,2$
 and $D=3,4,6,10$ one can construct a continuous family of inequivalent superstring actions,
 parameterized by the coefficient $c$ in front of the Wess-Zumino (WZ) term. For the standard GS
 choice $c=\pm 1$ the theory enjoys an additional fermionic gauge symmetry---the famous
 $\varkappa$-symmetry \cite{Siegel:1983hh}. For $\mathcal{N}=2$ and $c\neq\pm1$ we find that the
 classical integrability is lost, which could have been expected. However, somewhat surprisingly,
 for $\mathcal{N}=1$ we find that the theory is integrable for all values of $c$. On-shell
 amplitudes are independent of $c$, and the model exhibits an additional  hidden supersymmetry. In
 other words, $\mathcal{N}=1$ models without $\varkappa$-symmetry are equivalent to gauge fixed
 $\varkappa$-invariant $\mathcal{N}=2$ GS models. A similar observation for superparticles has been
 made recently in \cite{Mezincescu:2014zba}.

 In section~\ref{sec:conc} we conclude and discuss future directions.
\section{Current Algebra for Branes}
\label{sec:soft}
The main focus of our paper is to study the consistency of integrability with the non-linearly
realized target space Poincar\'e symmetry on the string worldsheet. However, it is instructive to
start with a few general observations on the implications of the target space Poincar\'e symmetry
for the brane worldvolume scattering.

There are two general methods to impose conditions of a non-linearly realized symmetry on scattering
amplitudes. First, one may follow the coset construction \cite{Callan:1969sn} to build the Goldstone
action invariant under the full symmetry group. Scattering amplitudes constructed from this action
are guaranteed to satisfy the required Ward identities. For non-linearly realized space-time
symmetries the corresponding construction has been worked out in
\cite{Isham:1971dv,Volkov:1973vd,Ivanov:1975zq}. At the leading order in the derivative expansion it
gives rise to the Nambu--Goto (NG) action for a $p$-brane,
\be
\label{braneNG}
S_{NG}=-\ell_s^{-p-1}\int d^{p+1}\sigma\sqrt{-\det\l{\eta_{\alpha\beta}+\ell_s^{p+1}\d_\alpha X^i\d_\beta X^i}\r}\;,
\ee
where $\sigma^\alpha$, $\alpha=0,\dots,p$ are the worldvolume coordinates, $\eta_{\alpha\beta}$ is
the worldvolume Lorentz metric, and $X^i$, $i=p+1,\dots, D$ are physical transverse excitations of
the brane (``branons"). Finally, $\ell_s$ is the brane width, so that the tension is equal to
$\ell_s^{-p-1}$.

This method is very physical and constructive, but it does not make immediately manifest special
properties of the scattering amplitudes  following from the non-linearly realized symmetry. For
example, the vanishing of the QCD scattering amplitudes in the limit when any of the external pion
legs become soft (a single soft pion theorem) is not immediate to see from the pion chiral
Lagrangian. The approach which allows one to directly see various soft theorems goes under the name of
current algebra (see, e.g., \cite{Bando:1987br} for a review).  The main idea is to systematically
study the Ward identities for the spontaneously broken currents.

Let us apply this method to brane worldvolume scattering and derive the analogue of the single
soft pion theorem in this case. Note that, unlike for the pion chiral Lagrangian, every Goldstone
field appears in (\ref{braneNG}) with a derivative acting on it. This is related to the trivial
commutation relations between spontaneously broken generators of translations and makes the
conventional single soft property of the branon scattering amplitudes obvious already at the level
of the action. However, one may expect more. Indeed, the conventional soft pion theorems reflect the
existence of a continuous family of vacua related by the action of broken symmetry generators (see,
e.g., \cite{ArkaniHamed:2008gz} for a recent discussion).  For a brane the moduli space is even
larger and includes the possibility of tilting the brane and moving it with a constant velocity,
{\it i.e.}, includes configurations like $X^i=v^i_\alpha \sigma^\alpha$. This property is also
related to the fact that one branon is responsible for a few broken symmetries---shifts, rotations
and boosts.  This gives rise to the expectation of an even stronger softness property for branon
amplitudes\footnote{Similar relations were recently observed between subleading soft theorems and
asymptotic symmetries in four-dimensional gravity \cite{Kapec:2014opa}.}. This intuition can be
converted into a more rigorous current algebra argument as follows.

Spontaneously broken translations of the target space give rise to conserved shift currents
$S_\alpha^i$, which take the form
\be
    \label{translations}
    S_\alpha^i=\ell_s^{1-p\over 2}\d_\alpha X^i+s_\alpha^i\;,
\ee
where  $s_\alpha^i$ is the non-linear part of the current.  In the same way as for pions, the
existence of these currents implies that the scattering amplitudes for the emission of a branon  are
soft, {\it i.e.}, they vanish in the limit $p^\alpha\to 0$, where $p^\alpha$ is the branon momentum.
However,  branon amplitudes have additional non-obvious softness properties following from the
existence of spontaneously broken boost currents $J_\alpha^{i\beta}$. The general form for these is
\be
    \label{boost}
    J_\alpha^{i\beta}=\ell_s^{-1}(\sigma^\beta S^i_\alpha-K_\alpha^{i\beta})\;,
\ee
where $K_\alpha^{i\beta}$ have no explicit dependence on the world-volume coordinates.  Conservation
of the boost currents implies that the shift current is itself a total derivative on shell,
\be
    \label{dboost}
    S^i_\alpha=\d_\beta K_\alpha^{i\beta}
\ee
or, equivalently,
\be
    \label{sia}
    s^i_\alpha=\d_\beta k_\alpha^{i\beta}\;,
\ee
where $ k_\alpha^{i\beta}$ is again the piece of $ K_\alpha^{i\beta}$ which depends non-linearly on
the fields.  To extract the consequences of non-linearly realized boosts one may either study Ward
identities involving boost currents, or make use of (\ref{dboost}) in the Ward identities
for the shift current itself. We will follow the latter path.

Note that it is straightforward to check  (\ref{dboost}) for the NG
action explicitly, and also to find the explicit form for the operator $K_\alpha^{i\beta}$ in this case.
Indeed, the only dependence of the NG action on the worldvolume Lorentz metric
$\eta_{\alpha\beta}$ is through the induced metric
\[
\gamma_{\alpha\beta}=\eta_{\alpha\beta}+\ell_s^{p+1}\d_\alpha X^i\d_\beta X^i\;.
\]
Hence, variation of the action under coordinate-dependent shifts
\[
X^i\to X^i+\ell_s^{1-p\over 2}\epsilon(\sigma)
\]
as required for deriving the Noether shift current, is equivalent to the
change of the metric according to
\[
\eta_{\alpha\beta}\to \eta_{\alpha\beta}+\ell_s^{3+p\over 2}(\d_\alpha\epsilon\d_\beta
X^i+\d_\beta\epsilon\d_\alpha X^i)\;.
\]
As a result one obtains
\be
    \label{SXT}
    S_\alpha^i=\ell_s^{3+p\over 2}\d_\beta X^i T^\beta_\alpha\;,
\ee
where $T^\beta_\alpha$ is the world-volume energy-momentum tensor. As a consequence of the conservation
of $T^\beta_\alpha$, the relation (\ref{dboost}) follows from (\ref{SXT}) on shell with
\be
\label{K}
    K_\alpha^{i\beta}=\ell_s^{3+p\over 2}X^i T^\beta_\alpha\;.
\ee
Let us now study the consequences of the non-linearly realized Poincar\'{e} symmetry by inspecting
matrix elements with a single insertion of the shift current in the form (\ref{dboost}).
In this case the reasoning goes very similar to the standard pion current algebra arguments. Namely,
one starts with the current conservation written in the form
\[
\langle out| \d^\alpha S^i_\alpha |in\rangle=0\;.
\]
By performing the Fourier transform and making use of (\ref{translations}), (\ref{sia}) we obtain
\be
    \label{dSp}
    \ell_s^{1-p\over 2} p^2\langle out|  X^i(p) |in\rangle+p^\alpha p_\beta\langle out|
    k_\alpha^{i\beta}(p) |in\rangle=0\;.
\ee
The matrix element entering the first term in (\ref{dSp}) has a one branon LSZ pole at $p^2=0$,
while the second matrix element is in general regular at $p^2=0$.  Consequently, restricting to the
case $p_0>0$ and taking the limit $p^2\to 0$, we obtain
\be
    \label{double_soft}
   i \ell_s^{1-p\over 2} \langle out|  in, p\rangle=-p^\alpha p_\beta\langle out|
    k_\alpha^{i\beta}(p) |in\rangle\;,
\ee
where $ |in, p\rangle$ is the initial state with an additional branon of momentum $p$.  By taking
now the soft limit $p^\alpha\to 0$ we conclude that non-linearly realized Poincar\'{e} symmetry
implies that scattering amplitudes with a single soft branon emission are {\it double} soft, {\it
i.e.} they vanish as a second power of the branon momentum.  Let us stress that the underlying
assumption for this argument is that matrix elements of $k_\alpha^{i\beta}$ are regular functions of
$p^\alpha$. For a general $p$-brane this should be the case, because $k_\alpha^{i\beta}$ is a
non-linear operator, however, we will see later that there is an interesting subtlety in the string
case, $p=1$.
\section{Bosonic Strings} %
\label{sec:bosonic}
Let us specialize now to the $p=1$ case, {\it i.e.} consider scattering on the worldsheet of a
string.  Let us first consider the case when the translational Goldstones $X^i$ are the only
massless degrees of freedom.  Their $SO(D-2)$-symmetric  $2\to2$ amplitude can be conveniently
parameterized in terms of three functions $A$, $B$, and $C$ called respectively annihilations,
transmissions and reflections,
\be
    \label{amplitude}
    {\cal M}_{ij,kl} = A \delta_{ij} \delta_{kl}+B \delta_{ik} \delta_{jl} + C \delta_{il} \delta_{jk}\:.
\ee
For $2\to 2$ scattering in two dimensions the momentum transfer between  particles is impossible,
hence one can always choose $p_1=-p_3$ and $p_2=-p_4$, which in terms of Mandelstam variables reads
\be
    -(p_1+p_2)^2=s=-u, \qquad t=0
\ee
Thus, effectively the amplitude is a function of only one variable. In what follows we consider
massless particles and it is convenient to distinguish left-movers with non-zero
$p_+\equiv{(p_0+p_1)/\sqrt{2}}$ and right-movers with non-zero $p_-\equiv{(p_0-p_1)/\sqrt{2}}$.
Crossing symmetry relates $A$, $B$ and $C$ as
\be
    A(s)=C(-s), \quad B(s)=B(-s).
    \label{crossing}
\ee
In particular, the absence of reflections is equivalent to the absence of annihilations.  At
tree-level the worldsheet $S$-matrix of the bosonic string is integrable and reflectionless in any
number of dimensions. A short proof of this statement goes as follows. Let us consider the
light-cone quantization of the bosonic string.  At the classical level the light-cone gauge is
perfectly consistent with all the symmetries, hence tree-level worldsheet amplitudes should be
correctly reproduced by this quantization for any number of space-time dimensions. The light-cone
energy levels depend only on the total left- and right-moving momenta of a state (``levels" of a
string state) and depend neither on how these momenta are distributed among individual
particles, nor on the $SO(D-2)$ quantum numbers of the state. This implies that there is no particle
production and that the scattering is reflectionless\footnote{See \cite{Dubovsky:2012wk} for a
detailed discussion of the connection between the light-cone energy spectrum and worldsheet
scattering amplitudes.}.  Then, to specify the full tree-level $S$-matrix, one only needs the
transmission part of the $2\to2$ amplitude, which reads
\be
    B_{tree}=\frac{\ell_s^2}{2}s^2 \ .
    \label{btree}
\ee
 Furthermore, the non-linearly realized Poincar\'{e} symmetry forbids any term in the action that
 can contribute at one-loop level to the on-shell $S$-matrix elements, thus also making the
 one-loop $S$-matrix universal and uniquely specified by symmetries.

The one-loop calculation of $2\to2$ scattering was performed in \cite{Dubovsky:2012sh}, where it was
found that the reflectionless property of the S-matrix is violated due to the presence of a finite
rational term in the amplitude\footnote{In the conformal gauge this term corresponds to the
Polchinski--Strominger operator \cite{Polchinski:1991ax}.}, unless the number of dimensions is equal
to 26,
\be
    A_{1-loop}=-C_{1-loop}=-\frac{D-26}{192\pi}\ell_s^4 s^3 \ .
    \label{1loop4pt}
\ee
Also the $D=3$ case is special, in this case there is only one flavor, making the notion of
annihilations meaningless, and the rational term in the amplitude vanishes as a consequence of
$A+C=0$.

\begin{figure}[t!]
    \begin{center}
        \includegraphics[height=3cm]{\diagram{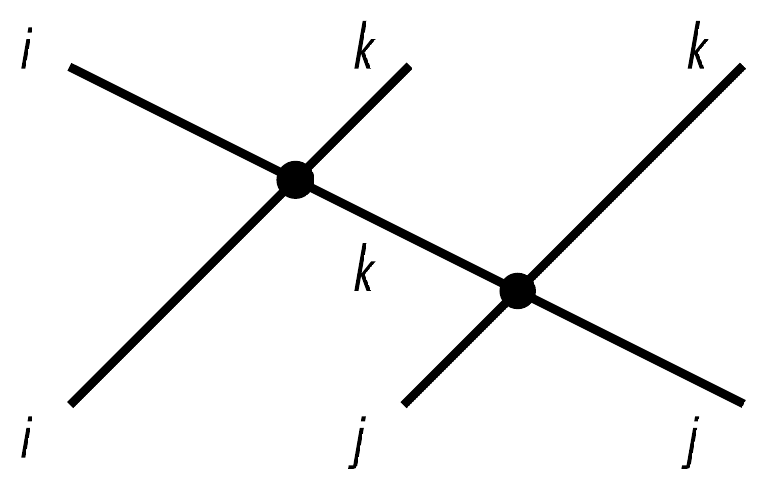}}
        \includegraphics[height=3cm]{\diagram{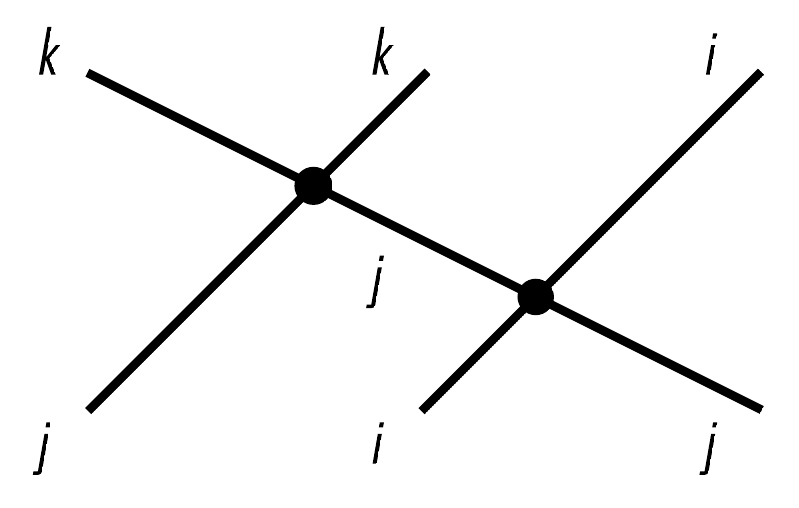}}
        \caption{Yang-Baxter equation cannot be satisfied in the $SO(N)$-symmetric massless theory in
        the presence of reflections for $N>2$}
        \label{fig:gs}
    \end{center}
\end{figure}

Even though the $2\to2$ scattering is always integrable in two dimensions due to purely kinematic
reasons, this result shows that the $S$-matrix cannot be integrable already at the level of $3\to3$
scattering in non-critical bulk dimensions larger than four. The reason is that  in an integrable
theory $3\to3$ $S$-matrix elements must satisfy the Yang-Baxter equation (see, e.g.,
\cite{Dorey:1996gd} for an introduction).  The Yang-Baxter equation can be understood as the
condition on an integrable $S$-matrix coming from the fact that the order in which the two-particle
scatterings occur does not affect the multi-particle $S$-matrix,
\be
S_{ij}^{kl}S_{km}^{np}=S_{im}^{kp}S_{kj}^{nl}.
\ee
When the number of flavors is larger than two it cannot be satisfied in the presence of reflections
and annihilations.{\footnote{Here we assume that there are no left-left and right-right scatterings,
which is true in our theory due to the absence of IR divergencies.}} To see this, note that for
three distinct flavors $i$, $j$ and $k$, the process $X^i(p_-) X^j(q_-) X^j(p_+) \to X^i(p_+)
X^k(p_-) X^k(q_-)$ (Figure \ref{fig:gs}, left) is possible, while the $X^j(q_-) X^i(p_-) X^j(p_+)$
state can only evolve into $X^k(p_+) X^k(q_-) X^i(p_-)$ state if one demands that $X^j$ particles
annihilate (Figure \ref{fig:gs}, right).

This argument, however, does not go through for two flavors. By considering all possible 3$\to$3
processes in $SO(2)$ symmetric theories we get the following condition from the Yang--Baxter
equation
\be
    A(s)=-C(s),
    \label{YB}
\ee
Note that the one-loop amplitude (\ref{1loop4pt}) satisfies this relation.

It is interesting to construct an example of an exact $S$-matrix satisfying this condition as well
as the usual requirements of unitarity and analyticity. These conditions are more naturally
formulated in terms of the $S$-matrix elements rather than amplitudes. We will denote them by
\be
a(s)=i\frac{A(s)}{2s}, \quad b(s)=1+i\frac{B(s)}{2s}, \quad c(s)=i\frac{C(s)}{2s},
\ee
where denominators are coming from the relativistic normalization for scattering states.  Then real
analyticity requires $a(s)^\dagger=a(s^\dagger)$ and the same for $b$ and $c$.  All these functions
have a cut going all the way along the real axis.  Finally, unitarity imposes the following
independent constraints
\be
bb^\dagger +aa^\dagger=1,\quad ba^\dagger+b^\dagger a=0,
\label{unitarity}
\ee

Combining real analyticity, (\ref{crossing}), (\ref{YB}) and (\ref{unitarity}) we get the following
equation for the ratio of $a$ and $b$
\be
\frac{a(s)}{b(s)}=\frac{a(-s)}{b(-s)}=\sqrt{\frac{b(s) b(-s)-1}{b(s)b(-s)}},
\ee
where we now understand $a$ and $b$ to be the meromorphic functions obtained by analytic
continuation from the physical sheet.

The simplest solution to this equation is
\bea
    b &=& \cos2\phi\frac{\l s-\frac{i\rho}{\sqrt{\cos 2 \phi}}\r^2}{\l s+i \rho \e^{i \phi}\r \l s+ i
    \rho\e^{-i\phi}\r}, \\
     a &=& i \sin 2\phi \frac{s^2 \l s-\frac{i\rho}{\sqrt{\cos 2 \phi}}\r^2}{ \l
    s+\frac{i\rho}{\sqrt{\cos 2 \phi}}\r \l s+i \rho \e^{i \phi}\r \l s+ i \rho\e^{-i\phi}\r}
    \label{ex}
\eea
In order to avoid poles on the physical sheet ({\it i.e.}, at ${\rm Im }\; s>0$) $\phi$ should be in
the interval $\phi\in[-\pi/4,\pi/4]$.  Other examples can be produced by multiplying this $S$-matrix
by CDD factors (cf.,\cite{Zamolodchikov:1978xm}).


To see whether there is a chance for a massless $O(2)$ invariant integrable theory of this kind to
exhibit the full four-dimensional Poincar\'e symmetry, let us inspect one-loop six-particle on-shell
amplitudes following from the NG action.  Just like one-loop four-particle amplitudes, these are
universal as long as branons are the only massless degrees of freedom.  Consequently, if some
one-loop six-particle amplitude does not vanish in non-integrable kinematics, integrability requires
extra gapless modes on a flux tube in four dimensions.

\begin{figure}[t!]
    \begin{center}
        \includegraphics[height=2.2cm]{\diagram{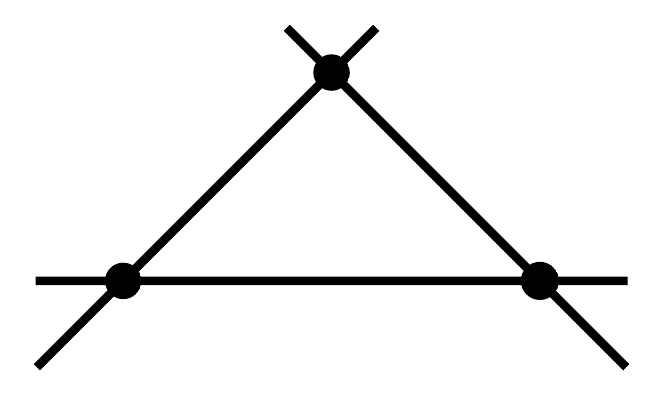}}
        \hspace{0.8cm}
        \includegraphics[height=2.1cm]{\diagram{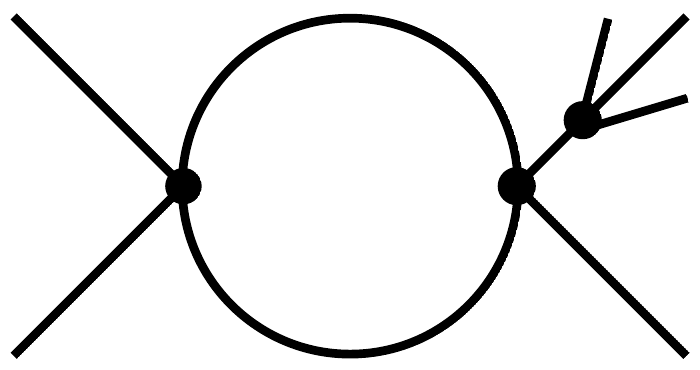}}
            \hspace{0.8cm}
          \includegraphics[height=2.1cm]{\diagram{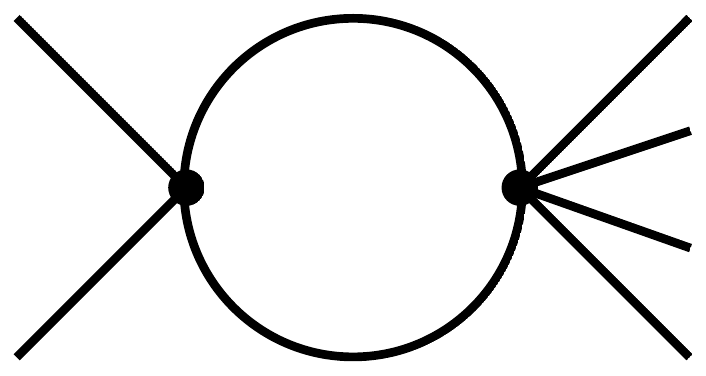}}
        \caption{Diagrams contributing to the one-loop branon 6-point function}
        \label{fig:diag}
    \end{center}
\end{figure}

It is a matter of a straightforward (even if a bit tedious) calculation to find out how these
amplitudes look.  We use dimensional regularization to preserve the non-linearly realized
Poincar\'{e} invariance.  As shown in Figure~\ref{fig:diag}, diagrams of three different topologies
contribute to the answer. For the sake of generality we explore the amplitude for a general target
space dimensionality.  There are two different types of kinematics allowed for the processes
involving six particles: two left-movers and four right-movers (or vice-versa) and three left-movers
and three right-movers.  We skip the details of the calculation, and only present the results here.
The most subtle part of this calculation is the treatment of the  Feynmann integrals corresponding
to the triangle graph in Figure~\ref{fig:diag}. We present the corresponding details in the
{\it Appendix}.

When the dust settles, it turns out that for non-integrable kinematics ({\it i.e.}, in the presence
of a non-trivial momentum  redistribution  between left- and/or right-movers) processes of the first
type have vanishing amplitudes.  However, we find that amplitudes with three right-movers and three
left-movers (which necessarily violate integrability) do not vanish, unless the string is critical,
$D=26$, or for $D=3$. To present the final answer in a compact form it is instructive to start with
a few general remarks about the possible structure of the amplitude.

The NG theory is integrable at the classical level. Moreover, for any $D$ one can construct an
integrable $S$-matrix, characterized by a two particle phase shift \cite{Dubovsky:2012wk},
\be
\label{GGRT}
\e^{2i\delta_{GGRT}(s)}=\e^ {i s\ell_s^2/4}\;,
\ee
which agrees with the NG theory at the classical level. Here GGRT stands for Goddard, Goldstone,
Rebbi and Thorn \cite{Goddard:1973qh}  because for any $D$ this $S$-matrix corresponds to the
light-cone quantization of the bosonic string. This implies that at the leading order in the
derivative expansion the difference between relativistic NG amplitudes and the GGRT amplitude
(\ref{GGRT}) should be a rational term, {\it i.e.} it should be possible to write a local vertex
reproducing the corresponding amplitude. Indeed, the $S$-matrix (\ref{GGRT}) satisfies all
analyticity, unitarity etc. requirements and weakly coupled at low energies. Hence,  one should be
able to write a local Lagrangian, which perturbatively reproduces it order-by-order in the
derivative expansion.  At the leading order this is the NG Lagrangian. If the latter gives a
different answer from (\ref{GGRT}) at a certain order, the difference can  be canceled by local
counterterms\footnote{Of course, in general, these counterterms will not respect nonlinearly
realized Poincar\'e symmetry.}.

For instance, the one-loop two particle annihilation amplitude discussed above can be reproduced
from the following local vertex,
\be
{\cal L}_{4}=\ell_s^4{D-26\over 48\pi}\d_+^2X^i\d_-^2X^i\d_+X^j\d_-X^j\;.
\ee
Then for the non-integrable part of the six particle amplitude we find that it corresponds to the
following local vertex,
\be
\label{PS6}
{\cal L}_{6}=\ell_s^6{D-26\over 48\pi}\d_+^2X^i\d_-^2X^i \l(\d_+X^j\d_-X^j)^2-{1\over 2}
   \d_+X^j\d_+X^j\d_-X^k\d_-X^k\r\;.
\ee

Both terms in (\ref{PS6}) encode a number of different processes  related by crossing symmetry and
violating integrability.  For instance, the first one gives rise to $2\to 4$ scattering of the
following flavor structure
\be
 X^i(p_++q_+) X^i(p_-+q_-)\to  X^j(p_+)X^j(p_-)X^k(q_+)X^k(q_-) \;,
    \label{state1}
\ee
the corresponding scattering amplitude is
\be
{\cal M}^{(1)}_{2\to 4}= \ell_s^6\frac{D-26}{24 \pi}p_- q_-(p_-+q_-)p_+ q_+
(q_++p_+)(p_+(2p_-+q_-)+q_+(2 q_-+p_-))\;.  \label{A1}
\ee
Diagrammatically, the only non-vanishing contribution to this process comes from the triangle graphs
in Figure~\ref{fig:diag}.

The second term in (\ref{PS6}) contributes, for example, into
\be
     X^i(p_++q_+) X^i(p_-+q_-)\to  X^j(p_+)X^j(q_+)X^k(p_-)X^k(q_-) \;.
    \label{state2}
\ee
In this case graphs of all three topologies contribute, and the result is
\be
{\cal M}^{(2)}_{2\to 4}= -\ell_s^6\frac{D-26}{24
    \pi} p_+q_+p_-q_-(p_++q_+)^2(p_-+q_-)^2\;.
    \label{A2}
\ee
For $D=4$ the effective non-integrable vertex (\ref{PS6}) simplifies and takes the following form
\be
 \label{PS64d}
 {\cal L}_{6,4d}=\ell_s^6{D-26\over 48\pi}\l\d_+^2{X^\dagger}\d_-^2X \l\d_-{X^\dagger}\d_+X\r^2 + h.c.\r\;,
 \ee
where $X=(X^1+iX^2)/\sqrt{2}$.  This completes the proof of the no-go theorem formulated in the {\it
Introduction}. The effective vertex (\ref{PS64d}) gives rise, for instance,  to the following
non-integrable process
\[
{X^\dagger}(p_++q_+)X(p_-+q_-)\to X^\dagger(p_+) X^\dagger(q_+) {X}(p_-){X}(q_-)\;,\] and the corresponding amplitude
is
\be
\label{A4d}
  {\cal M}^{4d}_{2\to 4}= \ell_s^6 \frac{11}{6\pi} p_-q_-p_+q_+(p_-+q_-)^2(p_++q_+)^2.
\ee
We see that if the branons are the only massless particles on a string, the classical integrability
is necessarily broken unless the string is critical, $D=26$, or there is a single branon, $D=3$, since in this case $X^\dagger=X$ and (\ref{PS64d}) vanishes on shell. In
these two special  cases we know the exact  $S$-matrix describing the corresponding integrable
theory, it is determined by the GGRT phase shift (\ref{GGRT}). Indeed, in both cases the light-cone
quantization is compatible with nonlinearly realized Lorentz symmetry. It remains to be seen whether
a consistent interacting $D=3$ string theory can be constructed. Even if this is the case, it
appears unlikely that it could be realized on the worldsheet of a confining string of some
conventional three-dimensional gauge theory. The reason is that the corresponding short strings
({\it i.e.}, strings with zero winding), which would correspond to glueballs, are anyons with
irrational spins \cite{Mezincescu:2010yp} at $D=3$. It would be surprising if one could obtain such
a spectrum in the large $N$ limit of some gauge theory, although it is definitely interesting to
understand whether the $D=3$ free string can be promoted into an interacting theory.


\subsection{Current Algebra for Strings}%
\label{CA}
A peculiar and somewhat surprising property of the amplitudes (\ref{A1}), (\ref{A2}) and (\ref{A4d})
is that they violate current algebra relations derived in Section~\ref{sec:soft}. Namely, they are
not double soft with respect to all external momenta.  What happens is that in two dimensions the
right hand side of (\ref{double_soft}) can develop a singularity when $p^\alpha$ goes to zero for
generic values of other momenta\footnote{Of course, also in higher dimensions singularities may
arise for special (collinear) kinematics. However, amplitudes remains double soft there for a {\it
generic} choice of momenta.}. The basic technical reason is that  collinear singularities may be
present for a generic kinematics in two dimensions.

Note that one might be surprised that we did not encounter IR singularities before. Indeed, the
standard lore says that  ``there are no Goldstone bosons in two dimensions" \cite{Coleman:1973ci},
and, more generally, massless two dimensional theories are plagued with IR singularities. However,
the string worldsheet theory provides a counterexample to this statement. We just saw that on-shell
worldsheet scattering amplitudes do not show any signs of IR divergencies, and the exact $S$-matrix
of a critical string (\ref{GGRT}) illustrates that there is no reason to expect IR divergencies at
higher loops.

Of course, technically the arguments presented in \cite{Coleman:1973ci} are all correct, however
they only prove that Goldstone fields $X^i$ do not give rise to well defined operators in a quantum
theory. In the string action there is always at least one derivative acting on every $X^i$.  This is
enough to ensure the absence of IR divergencies in on-shell scattering amplitudes.  On the other
hand, the non-linear part of the boost current (\ref{K}) contains  a Goldstone field without a
derivative acting on it, so it is not surprising that IR divergencies appear in the Ward identities
for this current.  In general, we expect that such singularities are related to lower order
amplitudes and matrix elements of some operators through unitarity, consequently the corrected Ward
identities and soft theorems can be constructed.

\begin{figure}[t!]
    \begin{center}
        \includegraphics[height=2.6cm]{\diagram{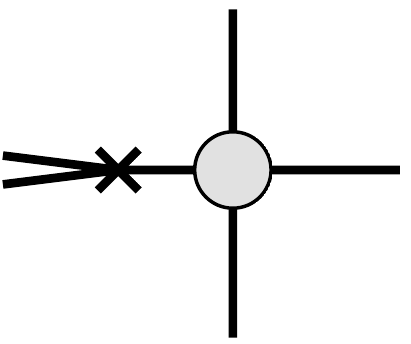}}
        \caption{A collinear singularity present in the shift current Ward identity}
        \label{fig:ward}
    \end{center}
\end{figure}

Let us illustrate now how this works using one-loop six particle amplitudes with all flavors $i,j,k$
different, which we already calculated by brute force, as an example. Then the most natural place
for collinear singularities to appear in (\ref{double_soft}) is through the diagrams of the type
shown in Figure~\ref{fig:ward}.  Indeed,  for concreteness, let us choose the soft momentum to be
left moving, $p_-=0$.  Then if all the external legs of the operator $k_\alpha^{i\beta}(p)$ are
right moving, the internal line connecting $k_\alpha^{i\beta}(p)$ to the rest of the diagram goes
on-shell in the soft limit, $p_+\to 0$, leading to an IR singularity.

Actually, one might worry that a collinear singularity could be present even before taking the soft
limit, if all the external legs of the operator  $k_\alpha^{i\beta}(p)$ are left moving. This would
invalidate the transition from the off-shell current conservation (\ref{dSp}) to the on-shell
relation (\ref{double_soft}). To see that this does not happen,  note that at the order we are
working we may  make use of (\ref{K}) and write
\be
\label{ppk}
p^\alpha p_\beta k_\alpha^{i\beta}=p_+^2X^i\tau_{--}+2p_+p_-X^i\tau_{+-}+p_-^2X^i\tau_{++}\;,
\ee
where $\tau_{\alpha\beta}$ is the energy-momentum tensor for $(D-2)$ free bosons.  In the on-shell
limit $p_-\to 0$ the collinear singularity from the internal line in Figure~\ref{fig:ward} scales as
$p_-^{-1}$. Then, given that $\tau_{+-}=0$ for free bosons, only the first term in (\ref{ppk})
survives on-shell.  Finally, for this term to contribute, at least one of the external legs attached
to it should be right moving.  We see that one can safely use  (\ref{double_soft}) and, moreover,
$p^\alpha p_\beta k_\alpha^{i\beta}(p)$ reduces just to $p_+^2 k_{--}^i$.
Furthermore in the soft limit, $p_+\to 0$, we can write
\be
    \langle out| k_{--}^{i}(p_+) |in\rangle =
     \sum_{j_1,j_2,k}
     i{\cal N}_3(p_{-j_1},p_{-j_2})
    \times\frac{i}{2 p_+(p_{-j_1}+p_{-j_2})}\times i {\cal M}_4((p_{-j_1}+p_{-j_2})_k)+reg.
    \label{sing}
\ee
Here the sum goes over all possible right-movers $j_1,j_2$ entering in $|in\rangle$, $|out\rangle$
and over all possible right-movers $k$, which combined with $j_1$, $j_2$, produce a non-vanishing
matrix element ${\cal N}_3(p_{-i},p_{-j})$ for the operator $ k_{--}^i(0)$. ${\cal
M}_4((p_{-j_1}+p_{-j_2})_k)$ is a four particle scattering amplitude between the state $k$ (carrying
the momentum $(p_{-j_1}+p_{-j_2})$) and the remaining particles in $|in\rangle$, $|out\rangle$.
Finally, $reg.$ stands for terms non-singular in the soft limit.

Singular terms in (\ref{sing}) imply violation of the double softness property of the amplitude.  We
see that Ward identities in this case do not imply that the amplitude is double soft, but rather
determine the single-soft piece of the amplitude through the amplitude with a smaller number of
particles and a three-particle matrix element of $ k_{--}^i(0)$. As follows from (\ref{K}), to
determine a singular piece we may replace this operator  by $\ell_s^2 X^i\d_- X^j \d_-X^j$. This is
true also for  amplitudes with a larger number of external states, since by Lorentz invariance the
parts of $k_{--}^i$ containing more than three fields will necessarily introduce  positive powers of
$p_+$.

Furthermore, for the kinematics with three left-movers and three right-movers, which we considered
in our brute force calculation, the most general effective vertex is a sum of two terms of the form
present in (\ref{PS6}).  Consequently, the functional form of the amplitude is determined up to two
numerical coefficients without need for any calculation.  Calculating the non-double soft part of
the amplitude through the relation (\ref{sing}) allows one to fix these coefficients and to determine
the full non-integrable part of the amplitude in a much more economic way, as compared to the brute
force calculation. For instance, one may take the soft $p_+\to 0$ limit of the scattering processes
(\ref{state1}) and (\ref{state2}).  In this case only the annihilation part (\ref{1loop4pt}) of the
4-point amplitude (\ref{amplitude}) contributes, and it is straightforward  to check that the
residue at the singularity in (\ref{sing}) agrees with the limit of (\ref{A1}) and (\ref{A2}) when
$p_+\to 0$.

It is worthwhile to push this program further and to study the subleading soft behavior of the world
sheet amplitudes when two or more momenta are taken to zero simultaneously. This limit should encode
the commutation relations of the bulk Poincar\'e algebra \cite{PhysRevLett.16.879} (see also
\cite{ArkaniHamed:2008gz}).
In particular, it will be interesting to see whether this method allows to prove that (\ref{GGRT})
in $D=3, 26$ is the only phase shift compatible with integrability and non-linearly realized
Poincar\'e symmetry (in the absence of additional massless particles).  This is not immediately
obvious, given that there are infinitely many integrable $S$-matrices given by pure CDD factors that
agree with (\ref{GGRT}) to the leading order in derivative and hence all share the same classical NG
action.  Our preliminary results indicate, however,  that the GGRT phase is indeed the only one
compatible with the Poncar\'e symmetry at the quantum level.

\begin{figure}[t!]
    \begin{center}
        \includegraphics[height=2.6cm]{\diagram{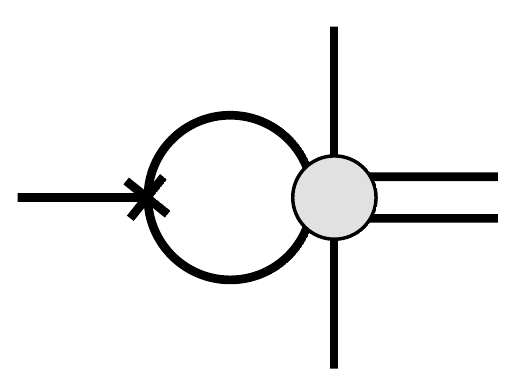}}
        \caption{Coleman-Thun type singularity in the shift current Ward identity}
        \label{fig:Coleman}
    \end{center}
\end{figure}

Note that, in general,  other types of singularities also contribute in (\ref{double_soft}). They
arise because loop diagrams leading to a cut in higher dimensions may result in a pole in 2D,
similarly to what happens in the sine-Gordon model \cite{Coleman:1978kk}. For example, the diagram
represented in Figure~\ref{fig:Coleman} will develop such a singularity when both momenta in the
loop become left-moving.
This does not happen at one loop for the processes with different flavors that we consider though,
because the shadowed part of the graph in Figure~\ref{fig:Coleman} vanishes as a consequence of
integrability and absence of annihilations at the tree level.


\section{Superstrings}
\label{sec:susy}
Let us see now whether the no-go result presented above can be avoided in the presence of additional
massless particles on the worldsheet. In this paper we restrict to the simplest possibility, namely
when the additional particles are all goldstini---massless fermions arising as a consequence of
non-linearly realized super-Poincar\'e symmetry.  In addition to the bosonic fields $X^i$ we now have
massless fermionic fields, $\theta^A$ which in what follows always belong to irreducible
representations of the bulk Lorentz group.
The Roman index $A$ runs from 1 to the total number of extended supersymmetries $\mathcal{N}$.
Usually the absence of higher spin particles imposes an upper bound on  $\mathcal{N}$, but from the
point of view of non-linear realization any $\mathcal{N}$ is consistent.

Let us first consider the situation when all bulk supersymmetries are broken by the
string\footnote{For constructions of brane actions partially breaking supersymmetry see {\it e.g.}
\cite{Hughes:1986dn},\cite{Ivanov:1999fwa}}.  Then the infinitesimal  SUSY transformations   are given by
\begin{equation}
\label{susyTransformation}
\delta \theta^A = \epsilon^A, \; \; \; \; \; \; \; \; \delta X^{\mu} = i \ell_s^2 \bar{\epsilon}^A \Gamma^{\mu}
\theta^A,
\end{equation}
where $\Gamma^{\mu}$ are D-dimensional gamma matrices.
Here we introduced
\[
X^\mu=(\sigma^\alpha, \ell_s X^i)
\]
so that SUSY transformations shift not only the fields, but also the worldsheet coordinates.

Generalizing the Nambu-Goto action to incorporate this symmetry and the new fermionic fields is done
in \cite{Hughes:1986dn} by promoting $\partial_{\alpha} X^{\mu}$ (which is trivially invariant under translations of the
bosonic embedding coordinates) to
\begin{equation}
\Pi_{\alpha}^{\mu} \equiv \partial_{\alpha} X^{\mu} - i \ell_s^2 \bar{\theta}^A \Gamma^{\mu} \partial_{\alpha}
\theta^A, \nonumber
\end{equation}
which is clearly invariant under super-translations as well.  Thus the action becomes
\begin{equation}
\label{SVA}
S_{VA} = - \frac{1}{\ell_s^2} \int d^2 \sigma \sqrt{-\mathrm{det} \Pi^{\mu}_{\alpha} \Pi_{\mu \beta}},
\end{equation}
which is a generalization of the action first considered by Volkov and Akulov
in \cite{Volkov:1972jx,Volkov:1973ix}.
 As first noticed by Green and Schwarz~\cite{Green:1983wt}, an additional term can be added to the
 Volkov-Akulov lagrangian (\ref{SVA}) at the same order in the derivative expansion for special
 values of $D$ and $\mathcal{N}$. This new term is SUSY invariant only up to a total derivative.
 Interpreting this theory as a non-linear sigma model parameterizing the coset supergroup
 $SISO(1,D-1)/SO(1,D-1)$, the origin of this term is the non-trivial Chevalley-Eilenberg cohomology
 of the corresponding Lie superalgebra \cite{Rabin:1985tv, Henneaux:1984mh}.  This term, which is
 analogous to the Wess-Zumino term of the pion lagrangian \cite{Wess:1971yu}, can only be written
 for $\mathcal{N}=1$ or $2$ and $D = 3, 4, 6$ and $10$ which will be referred to as ``special"
 dimensions\footnote{SUSY invariance of this term requires $\Gamma^{\mu} \theta_{[1}\bar{\theta}_2
 \Gamma_{\mu} \theta_{3]} = 0$ which holds when $D=3$, $\theta$ is Majorana; $D=4$, $\theta$ is
 Majorana or Weyl; $D=6$, $\theta$ is  Weyl, or $D=10$, $\theta$ is Majorana-Weyl.  Also, for Weyl
 spinors, every instance of $\bar{\theta} \Gamma^{\mu} \partial_{\alpha} \theta$ should be replaced
 by $(\bar{\theta} \Gamma^{\mu} \partial_{\alpha} \theta - \partial_{\alpha} \bar{\theta}
 \Gamma^{\mu} \theta)/2$.}.  It takes the following form
\begin{equation}
\label{wzAction}
S_{WZ} = i c \int d^2 \sigma \left( \epsilon^{\alpha \beta} \partial_{\alpha} X^{\mu}
(\bar{\theta^1} \Gamma_{\mu} \partial_{\beta} \theta^1 - \bar{\theta^2} \Gamma_{\mu}
\partial_{\beta} \theta^2) + i \ell_s^2 \bar{\theta^1} \Gamma^{\mu} \theta^1 \bar{\theta^2} \Gamma_{\mu}
\theta^2 \right).
\end{equation}
 The full action that we are going to study is then
\be
S=S_{VA}+S_{WZ}\;.
\label{action}
\ee
Unlike for the pion Lagrangian, the corresponding coupling constant $c$ does not have to be
quantized and can be arbitrary. The choice $c=\pm1$ corresponds to the GS superstring, which enjoys
a non-obvious fermonic gauge symmetry ($\varkappa$-symmetry).  In many respects, this case is
similar to the bosonic string. Namely, one can solve the theory exactly by using the light-cone
gauge quantization. Using the same arguments as in \cite{Dubovsky:2012wk}, one finds that the
resulting finite volume spectrum corresponds to integrable reflectionless scattering with $S$-matrix
determined by the phase shift (\ref{GGRT}). Famously, this quantization is consistent with
non-linearly realized Poincar\'e symmetry only at  $D=10$ (critical superstrings) and $D=3$
\cite{Mezincescu:2013xza} .

Here we first concentrate on the remaining ``non-special" choices of dimensions, and $c\neq\pm 1$
superstring in special dimensions. Later we will discuss the relation to the GS superstring as well.
The formulas below are presented for the $N=2$ superstring with a WZ term. To obtain the results for
the $N=1$ superstring, one  needs to set one of the spinors to zero. To obtain the results for
non-special dimensions, one needs to set $c=0$.

Since the unbroken linearly realized symmetry subgroup  is $SO(1,1) \times
SO(D-2)$, the following representation of the gamma matrices is convenient
\begin{align}
\Gamma^{\alpha} &= \rho^{\alpha}\otimes\mathbb{1} \nonumber \\
\Gamma^{j} &= \rho^{*}\otimes\gamma^{j}\;,
\label{2ind}
\end{align}
where $\gamma^{j}$ are $(D-2)$-dimensional gamma matrices, and $\rho^{\alpha}$ are two dimensional
gamma matrices, which  will be chosen  in the Weyl basis,
\begin{equation}
\rho^{0} = \left( \begin{array}{cc} 0 & -1 \\ 1 & 0 \end{array} \right), \; \rho^1 = \left(
\begin{array}{cc} 0 & 1 \\ 1 & 0 \end{array} \right), \; \rho^* = \rho^0 \rho^1. \nonumber
\end{equation}
\subsection{The Quartic Vertex}%

We expand the Lagrangian in $\ell_s$ and canonically
normalize the spinors as
\[
\theta^{A} \rightarrow \left( \l 2(\mathbb{1} + (-)^A c\rho^{*})\r^{-1/2}
\otimes\mathbb{1}\right)\theta^{A}\] to obtain
\begin{equation}
\mathcal{L} = - \tfrac{1}{2}\partial_{\alpha}X^{j}\partial^{\alpha}X_{j} + i\frac{1}{2}
\bar{\theta}^{A}\slashed{\partial}\theta^{A} + i\ell_s\frac{\left(\eta^{\alpha\beta} + (-)^A
c\epsilon^{\alpha\beta}\right)}{2\sqrt{1-c^{2}}}\left(
\bar{\theta}^{A}\rho^{*}\gamma^{j}\partial_{\alpha}\theta^{A} \right)\partial_{\beta}X_{j} + \ldots
\end{equation}
Since all cubic vertices vanish on-shell for any value of $c$ it is convenient to perform  the
following field redefinitions,  which remove them altogether,
\begin{align}
\label{remove3}
\theta^A \rightarrow & \theta^A +\ell_s \lambda_A \partial_{\alpha} X^i \rho^{*}\rho^{\alpha}\otimes
\gamma_{i} \theta^A \nonumber\\ X^i \rightarrow & X^i + \ell_s\eta_A \bar{\theta}^A \rho^{*} \otimes
\gamma^i \theta^{A}\;.
\end{align}
Here $A$ runs from 1 to 2 and summation is implied for the bosonic field redefinition. Notice that
 $X^{i}$ shifts by a pseudoscalar quantity, because the Wess-Zumino contribution
to the Lagrangian (\ref{wzAction}) is a pseudoscalar with respect to the worldsheet Lorentz group,
$SO(1,1)$.

The coefficients of the field redefinitions that cancel the three point vertices of the lagrangian
are universal ({\it i.e.}, the same for every $D$) and take the form
\begin{align}
\lambda_A = & - \frac{1 - (-)^A c}{\sqrt{1-c^2}} \nonumber \\
\eta_A = & \ \ \ (-)^A \frac{-c}{\sqrt{2-2c^2}} \nonumber
\end{align}

\begin{figure}[t!]
\centering
\begin{align*}
\frac{1}{4E_1E_2}\cdot \left(
\begin{array}{c}
\includegraphics[height=2.2cm]{\diagram{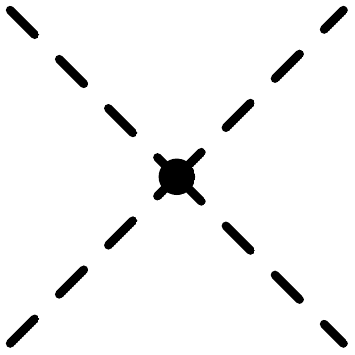}}
\end{array} \right) =  &
\frac{1}{2E_2}\cdot \left(
\begin{array}{c}
\includegraphics[height=2.2cm]{\diagram{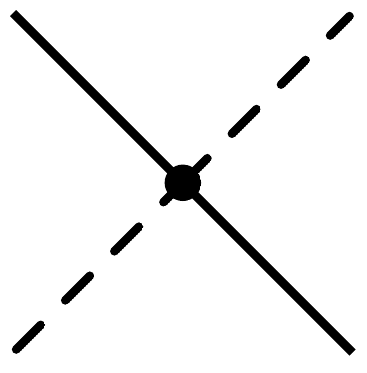}}
\end{array} \right) =
-1 \cdot\l\begin{array}{c}
\includegraphics[height=2.2cm]{\diagram{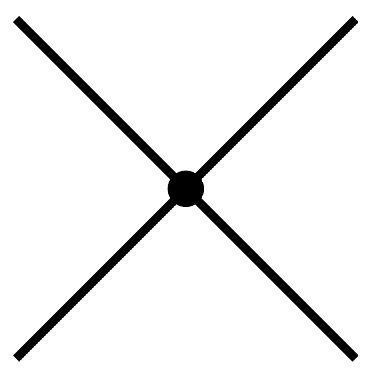}}
\end{array} \r=
\frac{i s \ell_s^2}{2} \ . \nonumber
\end{align*}
\caption{4-point interactions between branons and goldstini. The solid line represents a goldstino
corresponding to one broken supercharge. Time is flowing upwards (also in all later Figures).
The amplitudes containing bosons need to be divided by the norm of the bosonic states to obtain the
corresponding $S$-matrix element.} \label{5_diagrams}
\end{figure}

First, we restrict our attention to the case of $\mathcal{N}=1$ and special dimensions.
The results for the $2\to2$ scattering are presented in Figure~\ref{5_diagrams}.
Just like for a bosonic string, we find that the amplitude describes pure transmission. Moreover,
the corresponding phase shift is independent of $c$ and the same for all  processes (the relative
minus sign for the fermions comes from the ordering of in and out states, that is for a free fermion
the $S$-matrix would be $-1$). In other words, to the leading order the $S$-matrix is proportional
to the identity operator  and is given by (\ref{btree}), as before.

\begin{figure}
\centering
\begin{equation*}
\begin{array}{c}
\includegraphics[height=2.2cm]{\diagram{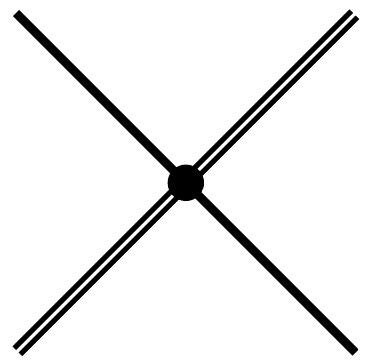}}
\end{array} = -i \frac{s \ell_s^2}{1+c} \ ,
\begin{array}{c}
\includegraphics[height=2.2cm]{\diagram{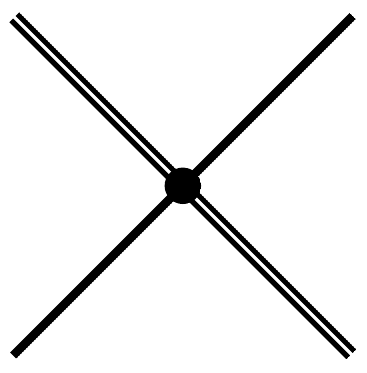}}
\end{array} = -i \frac{s \ell_s^2}{1-c} \ ,
\begin{array}{c}
\includegraphics[height=2.2cm]{\diagram{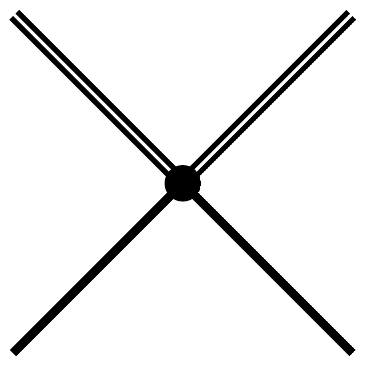}}
\end{array} = -i \frac{s \ell_s^2}{2} \ .
\end{equation*}
\caption{4-point interactions among fermions
corresponding to different broken supercharges, which are labeled by single and double lines.} \label{non_int}
\end{figure}

Let us now turn to the case of $\mathcal{N}=2$. Naturally, as long as the processes involving only
$A=1$ or $A=2$ fermions are considered, all properties of  $\mathcal{N}=1$ amplitudes remain.
However, processes involving both types of fermions are present, and for them the $S$-matrix is no
longer proportional to the identity.  It exhibits  both annihilations in the flavor $A$-space, and
non-trivial structure in $SO(D-2)$ spinor space.  Moreover, some of the amplitudes now depend on
$c$, and become divergent at $c = \pm 1$. Several examples of amplitudes (with  spinor structure
suppressed) are presented on Figure \ref{non_int}.

\subsection{The Quintic Vertex}%

From the previous experience with a  bosonic string, one may expect that a non-trivial structure of
the $2\to 2$ amplitudes indicates the breakdown of integrability for processes with a larger number
of particles. To check this, let us inspect five-particle processes. This is rather straightforward
because, after the field redefinition (\ref{remove3}), only the quintic vertices in the Lagrangian
contributes to this processes. One finds that the only quintic vertex which remains is the one with
four fermions and one boson.  For non-special number of dimensions it does not vanish on shell. The
same is true for ${\cal N}=2$, $c\neq\pm 1$, for processes involving fermions of different flavors.
This proves that superstrings are not integrable for these choices of the parameters.  However,
somewhat surprisingly, all quintic vertices vanish on shell for the ${\cal N}=1$ superstring in
special dimensions, independently of the value of $c$. We will discuss the reason and implications
of this cancellation in the next subsection. Before doing that, it is instructive to discuss in some
more detail the structure of the quintic vertex when it does not vanish.

\begin{figure}
\begin{equation*}
    \begin{array}{c}
    \includegraphics[height=2.2cm]{\diagram{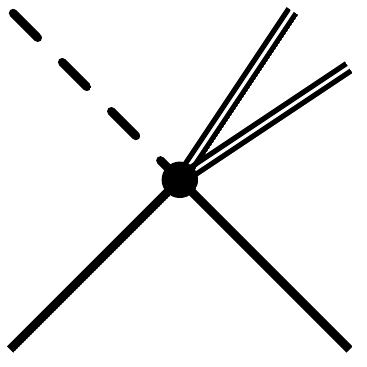}}
    \end{array} = i \ell_s^3\sqrt{\frac{1+c}{1-c}} p_+ p_-^2  \ .
\end{equation*}
\caption{The 5-point vertex does not vanish for $\mathcal{N}=2$ away from $c=\pm1$ for processes
involving both $A=1$ (single line) and $A=2$ (double line) goldstini.} \label{quintic}
\end{figure}

An example of a non-vanishing $\mathcal{N}=2$ quintic amplitude is shown in Figure~\ref{quintic},
where
$p_-$ is the momentum of one of the $A=2$ fermions, and we suppressed the spinor indices.  Values
$c=\pm1$ are special again due to the emergence of $\kappa$-symmetry at these points.
Even though all our fermions are goldstinos of spontaneously broken SUSY generators, the
non-vanishing quintic amplitudes are not soft with respect to the momenta of some of the fermions.
This implies that singular terms must be present in the SUSY Ward identities similarly to the
bosonic case of section \ref{CA}. These singularities, however, are not specific to 2 dimensions,
rather they are analogous to the contributions present in the case of pion-nucleon scattering and
related to the terms quadratic in fields present in the non-linear currents \cite{Bando:1987br}.

Indeed, after
canonically normalizing the fields, the SUSY transformations (\ref{susyTransformation}) become
\begin{align}
\delta_{\epsilon^{A}}\theta^{A'} &= \delta^{A'}_{A}    \left( \mathbb{1}+(-)^A c\rho^{*}
\right)^{+1/2}\epsilon^{A}    + i \ell_s \bar{\theta}^{A}\left(\mathbb{1}-(-)^A
c\rho^{*}\right)^{-1/2}\rho^{\alpha}\epsilon^{A}\partial_{\alpha}\theta^{A'}/2   \nonumber \\
\delta_{\epsilon^{A}}X^{j} &= -i \bar{\theta}^{A}\left(\mathbb{1}-(-)^A
c\rho^{*}\right)^{-1/2}\rho^{*}\gamma^{j}\epsilon^{A}/2 + i \ell_s \bar{\theta}^{A}\left(\mathbb{1}-(-)^A
c\rho^{*}\right)^{-1/2}\rho^{\alpha}\epsilon^{A}\partial_{\alpha}X^{j}/2 \ ,
\label{susytransformationsmodified}
\end{align}
and the conserved fermionic current corresponding to this symmetry is
\begin{equation}
\mathcal{J}^{\alpha A} = \left(\mathbb{1}-(-)^{A} c\rho^{*}\right)^{1/2}\rho^{\alpha}\theta^{A} +
\ell_s\left(\mathbb{1}+(-)^{A} c\rho^{*}\right)^{-1/2}\rho^{*}\gamma^{j}\theta^{A}\left(
\eta^{\alpha\beta}-(-)^{A} c\epsilon^{\alpha\beta}\right)\partial_{\beta}X_{j} + \ldots \ ,
\end{equation}
which gives the following Ward identity
\begin{equation}
\slashed{\partial}\left\langle out |\theta^{A} |in \right\rangle = -
\frac{\eta^{\alpha\beta}-(-)^{A} c\epsilon^{\alpha\beta}}{\sqrt{1-c^{2}}}\ell_s\left\langle out
|\rho^{*}\gamma^{j}\partial_{\alpha}\left(\theta^{A}\partial_{\beta}X_{j}\right)| in\right\rangle  +
\ldots  \ . \label{wardsusy}
\end{equation}

To see how this formula works in detail, we will consider a particular example in $\mathcal{N}=2$,
$D=4$. In the basis (\ref{2ind}) the Majorana fermion can be parametrized as
\be
\Theta^A=\l \l \begin{array}{c} \alpha_A \\ \alpha_A^\dagger \end{array} \r , \l \begin{array}{c}
\beta_A \\ \beta_A^\dagger \end{array} \r \r
\ee
where the $\alpha$'s are right-moving and $\beta$'s are left-moving, and both have positive $SO(2)$
helicities. In this notation the kinetic term reads
\be
-i\sqrt{2}\alpha_A \d_+\alpha_A^\dagger-i\sqrt{2}\beta_A \d_-\beta^\dagger_A \;.
\ee
Since the quintic vertex that we consider has three derivatives, the worldsheet Lorentz invariance
requires that the process involving, say, a left-moving boson always has one left-moving and three
right-moving fermions.

\begin{figure}
\begin{equation*}
    \begin{array}{c}
    \includegraphics[height=2.2cm]{\diagram{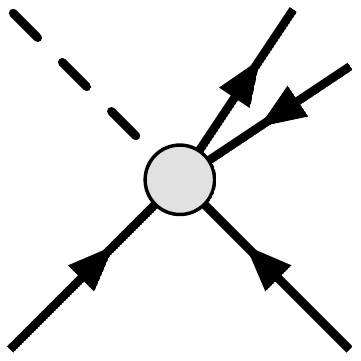}}
    \end{array} =
    \begin{array}{c}
    \includegraphics[height=2.2cm]{\diagram{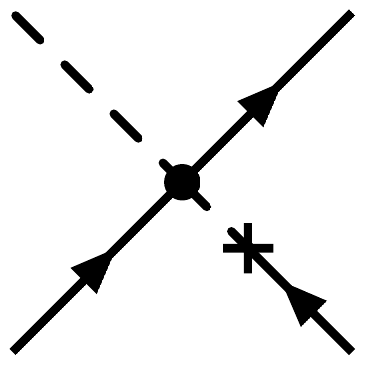}}
    \end{array} +
    \begin{array}{c}
    \includegraphics[height=2.2cm]{\diagram{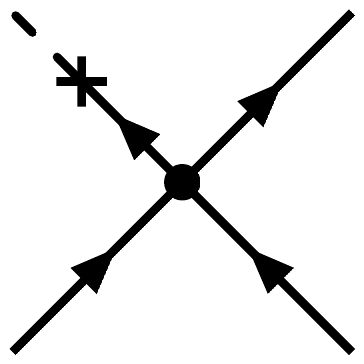}}
    \end{array} + \ldots
\end{equation*}
\caption{Singular diagrams in Ward identities for a 5-particle process involving one branon and
goldstini of one supercharge.  Arrows show the flow of the $SO(2)$ helicity.}
\label{diag::5ward1bosN1}
\end{figure}

First, let us consider a process where all fermions are of the same flavor and demonstrate that it
has a vanishing amplitude as a consequence of Ward identities. To be specific, let us take the
following initial and final states:
\be
\alpha_{1}(p_{-})\beta_{1}(p_{+})\rightarrow
\alpha_{1}^\dagger(k_{-})\alpha_{1}(p_- \!-k_{-})X(p_{+})
\label{f1} \ .
\ee
The only vertex that can contribute to this amplitude is
\be
\label{Q1ver}
\d_-\alpha_1 \alpha_1 \alpha_1^\dagger\d_+\beta \d_+X^\dagger\;.
\ee
The corresponding amplitude is not soft with respect to both $k_-$ and $q_-=p_- \!-k_{-}$. Let us
check whether this is consistent with the Ward identity (\ref{wardsusy}) applied to the process
(\ref{f1}) in the limit $k_-\to0$. The diagrams that can develop a singularity are shown in
Figure~\ref{diag::5ward1bosN1}.
In the soft limit, the internal propagators of both diagrams go on shell, and the matrix element
factorizes into the product of the 4-point amplitude, the propagator, and the matrix element of the
operator in the right hand side of (\ref{wardsusy}) between a boson and a fermion analogously to
(\ref{sing}).
%
Since we found that the corresponding $S$-matrices are equal in magnitudes but have opposite signs
(see Figure \ref{5_diagrams}) the sum of these two diagrams vanish. We conclude then that the vertex
(\ref{Q1ver}) should vanish, in agreement with what we found by a brute force calculation.

\begin{figure}
\begin{align*}
    \begin{array}{c}
    \includegraphics[height=2.2cm]{\diagram{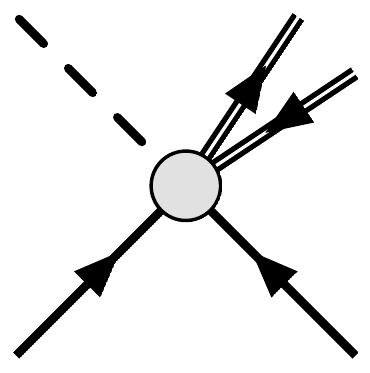}}
    \end{array} & =
    \begin{array}{c}
    \includegraphics[height=2.2cm]{\diagram{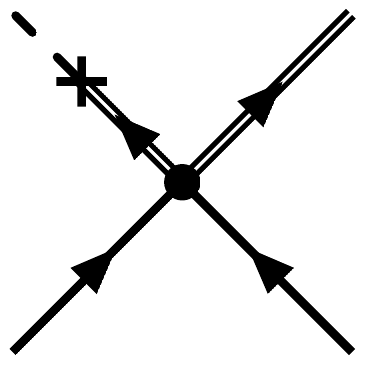}}
    \end{array} + \ldots                                      \\
\end{align*}
\caption{Singular diagram in Ward identities for a 5-particle process involving one boson and two different flavors of
fermions.} \label{diag::5ward1bosN2}
\end{figure}

On the other hand, let us consider the process
\be
\alpha_{1}(p_{-})\beta_1(p_+)\rightarrow
\alpha_{2}^\dagger(k_{-})\alpha_{2}(q_{-})X^\dagger(p_{+}) \ .
\ee
As a result of our field redefinition, there is a single vertex which contributes to this process,
which is
\be
\ell_s^3\sqrt{\frac{1+c}{1-c}}\d_-\alpha_2^\dagger \alpha_2 \alpha_1\d_+\beta_1\d_+X^\dagger \ ,
\ee
and the corresponding amplitude is
\be
\mathcal{M}_{(5)}=i\ell_s^3\sqrt{\frac{1+c}{1-c}}q_-p_+^2\;,
\ee
which is not soft with respect to $k_-$.  To see whether this behavior is consistent with the Ward
identities we again inspect the right-hand side  of (\ref{wardsusy}) in the limit $k_{-}\rightarrow
0$. This time there is only one singular contribution as shown in Figure \ref{diag::5ward1bosN2},

\be
\mathcal{M}_{(5)}= k_{-}\left(
-p_+\frac{1+c}{\sqrt{1-c^{2}}}\ell_s \right) \frac{-i}{k_{-}}  i\mathcal{M}_{(4)}\left(s\right)+O(k_-)
\ee
where $s=2 p_{-}p_{+}$ and  $\mathcal{M}_{(4)}$ is the third
amplitude in Figure~\ref{non_int}. This illustrates that the Ward identities indeed fix the quintic amplitude.

Note that $\mathcal{M}_{(5)}$ is soft with respect to $q_-$. It also follows immediately from the
Ward identity argument since the singular contribution in the $q_-\to0$ limit would require the
$\alpha_1\beta_1\to \alpha_2^\dagger \beta_2^\dagger$ part of $\mathcal{M}_{(4)}$ which is zero due
to  $SO(2)$ helicity conservation.


To summarize this section, we  see that in addition to the conventional GS superstring, only
$\mathcal{N}=1$ theories in $D=3,4,6,10$ with arbitrary value of $c$ stay as
viable candidates  to be integrable. Let us take now a closer look at this remaining set of models.


\subsection{$\mathcal{N} = 1$: A Closer Look}%
Interestingly, all $\mathcal{N} = 1$,  $D=3,4,6,10$ amplitudes,  which we calculated so far, are
independent of the Wess--Zumino coupling constant $c$.  Furthermore, the set of physical degrees of
freedom in these theories is equivalent to the $\mathcal{N}=2$ Green-Schwarz superstring after all
gauge degrees of freedom are removed. This suggests that these models are all equivalent to
$\mathcal{N} = 2$ GS superstrings at the same number of dimensions. In fact, this observation has
already been made for $c=0$ \cite{1986PhLB..182...33K} (see also \cite{Siegel:1983ke} for a detailed
analysis of the $D=3$, $c=0$ case). In this case, one can start with $\mathcal{N}=2$ GS superstring and
fix the $\varkappa$-symmetry, by imposing the condition $\theta^1=\theta^2$. The resulting action is
$\mathcal{N}=1$, $c=0$ superstring, which proves the equivalence at $c=0$. To extend the prove for
all values of $c\neq\pm 1$ we will show now that the WZ term can be removed by a field redefinition,
and hence  the theories are indeed equivalent for all values of $c$.

To do this, we need to look in some detail at the full classical field  equations for
$\mathcal{N}=1$ theory.  For this purpose it is convenient to introduce $X^0$ and $X^1$ as
dynamical fields and rewrite the action in the Polyakov form
\begin{equation}
\label{nisoneAction}
S = \int d^2 \sigma  \left( -\frac{1}{2}  \sqrt{-h} h^{\alpha \beta} \Pi^{\mu}_{\alpha} \Pi_{\beta
\mu} - i c \epsilon^{\alpha \beta} \bar{\theta} \hat{\Pi}_{\alpha} \partial_{\beta} \theta \right)
\end{equation}
where $\ell_s$ is set to unity since it is no longer needed as an expansion parameter and
 $\hat{\Pi}_{\alpha} \equiv \Gamma_{\mu} \Pi^{\mu}_{\alpha}$. Here $\theta$ satisfies the following identities
\begin{gather}
\label{fierz1} \bar{\theta}_1 \Gamma^{\mu} \theta_2 = - \bar{\theta}_2 \Gamma^{\mu} \theta_1 \\
\label{fierz2} \Gamma^{\mu} \theta_{[1} \bar{\theta}_{2} \Gamma_{\mu} \theta_{3]} = 0\;,
\end{gather}
which are specific to the GS choices of $D$ and $\theta$-representations, and ensure that the WZ
term is invariant under SUSY transformations.  At $c\not= \pm1$ this action has no local fermonic
symmetry and explicitly has only $\mathcal{N} = 1$ supersymmetry.

 Let us show now that for any $c\neq \pm 1$ the WZ term in the action vanishes on shell ({\it i.e.},
 as a consequence of the exact field equations). This will imply, that the parameter $c\neq\pm 1$
 can indeed be changed arbitrarily by a field redefinition and that this whole one parameter family
 of theories is equivalent to the $\mathcal{N} = 2$ Green-Schwarz action at $c= \pm 1$.

The variation of the action with respect to the auxiliary metric yields the following constraints
\begin{equation}
\label{hEom}T_{\alpha \beta} \equiv -\frac{2}{\sqrt{-h}} \frac{\delta S}{\delta h^{\alpha \beta}} =
\Pi^{\mu}_{\alpha} \Pi_{\beta \mu} - \frac{1}{2} h_{\alpha \beta} h^{\gamma \delta}
\Pi^{\mu}_{\gamma} \Pi_{\delta \mu}.
\end{equation}
Note that as a consequence of Weyl invariance the energy momentum tensor is traceless.

The Euler--Lagrange equations for scalars and fermions are
\begin{gather}
\label{XEom} \frac{\delta S}{\delta X^{\mu}} = \partial_{\alpha} \left( \sqrt{-h} h^{\alpha \beta}
\Pi_{\beta \mu} + i c \epsilon^{\alpha \beta} \bar{\theta} \Gamma_{\mu} \partial_{\beta} \theta
\right)=0
 \\
 \label{thetaEom} \frac{1}{2} \frac{\delta S}{\delta \bar{\theta}} = i(\sqrt{-h} h^{\alpha
\beta} - c \epsilon^{\alpha \beta}) \hat{\Pi}_{\alpha} \partial_{\beta} \theta + \frac{i}{2}
\partial_{\alpha} \left( \sqrt{-h} h^{\alpha \beta} \hat{\Pi}_{\beta} \right) \theta + c
\epsilon^{\alpha \beta} \Gamma^{\mu} \partial_{\beta} \theta \bar{\theta} \Gamma_{\mu}
\partial_{\alpha} \theta = 0.
\end{gather}
The Fierz identities (\ref{fierz1}) and (\ref{fierz2}) imply
\begin{equation}
\epsilon^{\alpha \beta} \Gamma^{\mu} \partial_{\beta} \theta (\bar{\theta} \Gamma_{\mu}
\partial_{\alpha} \theta ) = -\frac{1}{2} \epsilon^{\alpha \beta} \Gamma^{\mu} \theta
(\partial_{\alpha} \bar{\theta} \Gamma_{\mu} \partial_{\beta} \theta ), \nonumber
\end{equation}
which allows one to show that the last two terms of (\ref{thetaEom})
vanish due to the scalar field equations (\ref{XEom}). To further simplify the field equations,
we choose the orthogonal gauge
\begin{equation}
h^{\alpha \beta} \sim \eta^{\alpha \beta} =
\left( \begin{array}{cc}
0 & 1 \\
1 & 0 \end{array} \right). \nonumber
\end{equation}
In this gauge, labeling the coordinates as $``+"$ and $``-"$,  the field equations become
\begin{gather}
\label{fe1} \Pi_+^2 = \Pi_-^2 = 0 \\
\label{fe2} (1+c)\partial_- \Pi_+ + (1-c) \partial_+ \Pi_- = 0 \\
\label{fe3} (1+c) \hat{\Pi}_- \partial_+ \theta + (1-c) \hat{\Pi}_+ \partial_- \theta = 0.
\end{gather}
Note that as a consequence of (\ref{fe1}) the operators
\begin{equation}
\hat{\Pi}_{\pm} \equiv \Gamma^{\mu} \Pi_{\mu \pm}, \nonumber
\end{equation}
are nilpotent, $\hat{\Pi}^2_{\pm} = 0$. Then from (\ref{fe3}) it follows that
\be
\hat{\Pi}_+\hat{\Pi}_- \partial_+ \theta=\hat{\Pi}_- \hat{\Pi}_+ \partial_- \theta = 0\;.
\label{P}
\ee
We will restrict ourselves to configurations in which the
operator $\hat{P} \equiv \hat{\Pi}_- + \hat{\Pi}_+$ has no null vectors\footnote{This operator has a
term that looks like $\partial_0 X^{\mu} \Gamma_{\mu}$ which has a component proportional
to $\partial_0 X^0$. On a long string background in the static gauge, this is simply 1. Therefore,
working in effective field theory with small transverse fluctuations, this operator should be in the
vicinity of the identity.}. Then (\ref{P}) implies that
\begin{equation}
\hat{P} (\hat{\Pi}_- \partial_+ \theta) = \hat{P} (\hat{\Pi}_+ \partial_- \theta)  = 0, \nonumber
\end{equation}
hence
\begin{equation}
\label{nullCondition}
\hat{\Pi}_- \partial_+ \theta =
\hat{\Pi}_+ \partial_- \theta = 0,
\end{equation}
on the equations of motion for any $c \neq \pm1$.  This implies that the WZ action, which can be written as
\begin{equation}
S_{WZ} = i c \int d^2 \sigma (\bar{\theta} \hat{\Pi}_+ \partial_- \theta - \bar{\theta} \hat{\Pi}_-
\partial_+ \theta), \nonumber
\end{equation}
vanishes as a consequence of field equations for all $c\neq \pm1$, so that this term can be removed
by a field redefinition and theories with different values of $c$ are equivalent.  At the end of the
next section we will present an explicit form for the corresponding infinitesimal field redefinition
connecting different values of $c$.

\subsection{Hidden Symmetry}%

The arguments above explain why $c \neq \pm1$, $\mathcal{N} = 1$ theories are all equivalent to
$\mathcal{N} = 2$ GS superstrings.  This equivalence implies that $\mathcal{N} = 1$ theories should
have an additional hidden supersymmetry, corresponding to the second supercharge of the $\mathcal{N}
= 2$ GS theory. This supersymmetry should be realized linearly.  Given that  $\mathcal{N} = 1$
theories have enhanced infinite dimensional fermionic symmetries  at $c=\pm 1$
($\varkappa$-symmetries), to identify the hidden symmetry it is natural to inspect whether any
combination of $\varkappa$-symmetry transformations survives at $c\neq \pm 1$ as well.  The most
general combination of two infinitesimal  $\varkappa$-symmetry   transformations can be written as
\begin{gather}
\delta \theta = \hat{\Pi}_{\alpha} (e^{\alpha}_- \kappa_+ + e^{\alpha}_+ \kappa_-)
\nonumber \\ \delta X^{\mu} = i \bar{\theta} \Gamma^{\mu} \delta \theta \nonumber \\ \delta
\Pi^{\mu}_{\alpha} = 2 i \partial_{\alpha} \bar{\theta} \Gamma^{\mu} \delta \theta \nonumber \\
\delta e^{\alpha}_+ = 4 i e^{\alpha}_- e^{\beta}_+ \bar{\kappa}_- \partial_{\beta} \theta \nonumber
\\ \delta e^{\alpha}_- = 4 i e^{\alpha}_+ e^{\beta}_- \bar{\kappa}_+ \partial_{\beta} \theta
\nonumber \\ \label{kappaTransformation} \delta \left( \sqrt{-h} h^{\alpha \beta} \right) = 8 i
\sqrt{-h} (e^{\alpha}_- e^{\gamma}_+ e^{\beta}_- \bar{\kappa}_+ + e^{\alpha}_+ e^{\gamma}_-
e^{\beta}_+ \bar{\kappa}_-) \partial_{\gamma} \theta\;,
\end{gather}
where we introduced the ``zweibein" fields, $e_\pm^\alpha$, such that $h^{\alpha\beta}=e_+^\alpha
e_-^\beta+ e_+^\beta e_-^\alpha $.  One can check that these transformations leave the action
invariant when one picks $c=\pm1, \kappa_{\mp}=0$ and arbitrary $\kappa_{\pm}$.  For a general $c$,
the action variation under  a general combination of $\kappa$-transformations takes the following
form
 \begin{equation}
\label{kappaVariation}
\delta S = - 2 i \int d^2 \sigma \epsilon^{\alpha \beta} \partial_{\alpha} \bar{\theta}
\hat{\Pi}_{\beta} \left( (c-1) \hat{\Pi}_- \kappa_+ + (c+1) \hat{\Pi}_+ \kappa_- \right),
\end{equation}
where operators $\hat{\Pi}_{\pm} \equiv \hat{\Pi}_{\alpha} e^{\alpha}_{\pm}$ coincide in the
conformal gauge with $\hat{\Pi}_{\pm}$ used in the previous section.  For this variation to be zero,
we need to chose spinors $\kappa_{\pm}$ in such a way that
\begin{equation}
\label{strongCondition}
(c-1) \hat{\Pi}_- \kappa_+ + (c+1) \hat{\Pi}_+ \kappa_- = 0.
\end{equation}
For a generic off-shell configuration $\hat{\Pi}_{\pm}$ are non-degenerate, so one can use
(\ref{strongCondition}) to express $\kappa_+$ through $\kappa_-$ and naively one is left with a
local symmetry for any $c$.  However, for all on-shell configurations $\hat{\Pi}_{\pm}$ are
nilpotent
and the only way to solve (\ref{strongCondition}) is  to set separetly
\begin{equation}
\hat{\Pi}_- \kappa_+ =  \hat{\Pi}_+ \kappa_- = 0.
\end{equation}
Unfortunately, transformations of this form act trivially on-shell and do not lead to any
interesting symmetry. The way out is to note the following consequence of the Fierz identities
(\ref{fe1}) and (\ref{fe2}),
\begin{equation}
\epsilon^{\alpha \beta} \Gamma^{\mu} \theta ( \partial_{\alpha} \bar{\theta} \Gamma_{\mu}
\partial_{\beta} \theta) = \frac{2}{3} \epsilon^{\alpha \beta} \partial_{\alpha} (\Gamma^{\mu}
\theta (\bar{\theta} \Gamma_{\mu} \partial_{\beta} \theta)), \nonumber
\end{equation}
implying that $\epsilon^{\alpha \beta} \partial_{\alpha} \bar{\theta} \hat{\Pi}_{\beta}$ is a
total derivative,
\begin{equation}
\epsilon^{\alpha \beta} \partial_{\alpha} \bar{\theta} \hat{\Pi}_{\beta} = \epsilon^{\alpha \beta}
\partial_{\alpha} \left( \bar{\theta} \Gamma_{\mu} \left( \partial_{\beta} X^{\mu} - \frac{i}{3}
\bar{\theta} \Gamma^{\mu} \partial_{\beta} \theta \right) \right). \nonumber
\end{equation}
This means that we can integrate (\ref{kappaVariation}) by parts and the variation of the action
will vanish under a weaker condition, as compared to (\ref{strongCondition}),
\begin{equation}
(c-1) \hat{\Pi}_- \kappa_+ + (c+1) \hat{\Pi}_+ \kappa_- = const. \nonumber
\end{equation}
This weaker condition allows us to solve for a new non-trivial symmetry of the action, namely
\begin{equation}
\label{hiddenSymmetry}
\kappa_{\pm} = \frac{\hat{\Pi}_{\pm}}{2 (c\mp1) \Pi^{\mu}_+ \Pi_{-\mu}} \kappa,
\end{equation}
where $\kappa$ is a constant spinor. On a long string background this generator is spontaneously
broken.  However, one can choose a linear combination of the hidden SUSY transformation and of the
conventional one (\ref{susyTransformation}) which is realized linearly.

To complete the discussion of this new symmetry, we derive the associated conserved N\"{o}ether
current via the usual prescription.  Take the parameter of the transformation $\kappa$ to be an
arbitrary function on the worldsheet.  Then the variation of the action (\ref{nisoneAction}) takes
the following form
\begin{equation}
\delta S = - 2i \int d^2 \sigma \epsilon^{\alpha \beta} \partial_{\alpha} \bar{\kappa} \left(
\partial_{\beta} X^{\mu} - \frac{i}{3} \bar{\theta} \Gamma^{\mu} \partial_{\beta} \theta \right)
\Gamma_{\mu} \theta, \nonumber
\end{equation}
which yields the supercurrent
\begin{equation}
j^{\alpha}_{\kappa} = 2 i \frac{\epsilon^{\alpha \beta}}{\sqrt{-h}} \left( \partial_{\beta} X^{\mu}
- \frac{i}{3} \bar{\theta} \Gamma^{\mu} \partial_{\beta} \theta \right) \Gamma_{\mu} \theta.
\nonumber
\end{equation}
On the other hand, the conventional supercurrent corresponding to transformations
(\ref{susyTransformation}) is
\begin{equation}
j^{\alpha}_{\epsilon} = 2 i  h^{\alpha \beta} \hat{\Pi}_{\beta} \theta + 2 i c
\frac{\epsilon^{\alpha \beta}}{\sqrt{-h}} \left( \partial_{\beta} X^{\mu} + \frac{i}{3} \bar{\theta}
\Gamma^{\mu} \partial_{\beta} \theta \right) \Gamma_{\mu} \theta. \nonumber
\end{equation}
The existence of a hidden linearly realized SUSY confirms the equivalence of the model to the
standard $\mathcal{N}=2$ GS supersting.  Since both currents have the same chirality our
construction corresponds to the chiral IIB theory.

Finally, to present the field redefinition, which shifts the coefficient of the WZ term in the
action, let us modify the hidden symmetry (\ref{hiddenSymmetry}) to
\begin{equation}
\label{cshift}
\kappa_{\pm} = \epsilon \frac{\hat{\Pi}_{\pm}}{4 (c \mp 1) \Pi^{\mu}_+ \Pi_{- \mu}} \theta,
\end{equation}
where $\epsilon$ is some infinitesimal c-number. When these choices are made for the infinitesimal
spinors in a general $\kappa$ transformation (\ref{kappaTransformation}), from
(\ref{kappaVariation}) it is straightforward to demonstrate that
\begin{equation}
\delta_{\epsilon} S = i \epsilon \int (\bar{\theta} \hat{\Pi}_+ \partial_- \theta - \bar{\theta}
\hat{\Pi}_- \partial_+ \theta) = \epsilon S_{WZ},
\end{equation}
which is exactly what required for
a field redefinition which  changes
the value of $c$.
%
%
%
\section{Conclusions} \label{sec:conc}%

To summarize, we feel that the presented analysis elucidates the relation between integrability on
the fluxtube worldsheet and the conventional condition for the string to be critical. As our results
show, if the only gapless degrees of freedom on the flux tube are Goldstones and Goldstini of
spontaneously broken flat space-time (super)symmetries, then in addition to conventional critical
strings, the only other case when the worldsheet theory may be integrable is $D=3$.  It will be
interesting to understand whether these $D=3$ theories (which were studied recently in
\cite{Mezincescu:2010yp,Mezincescu:2011nh}) may be obtained from a consistent interacting bulk
theory. One puzzle related to them is that $D=3$ case does not appear special neither in the
Polchinski--Strominger description, nor  in the conventional Liouville approach to non-critical
strings.

Coming back to the initial question which motivated our analysis, namely the search for integrable
confining strings in four dimensional large $N$ gauge theories, we see that theories of this kind
necessarily need to carry additional gapless degrees of freedom.  Of course, this still leaves a
room for many interesting possibilities. It appears likely that answering this question will require
using more sophisticated field/string theory tools.  The approach of this paper, based on
universality of worldsheet scattering at (several) leading orders in the derivative expansion, will
hopefully serve as a useful cross-check and/or guideline in this search.

For example, the integrability condition can be satisfied by adding massless fermions, which are not
related to spontaneously broken supersymmetry.  We checked, for instance, the behavior of the
one-loop bosonic $2\to2$ amplitude in the presence of $N$ worldsheet Dirac (or $4N$ Majorana-Weyl)
fermions, that trivially transform under $SO(D-2)$. The result is equal to (\ref{1loop4pt}) with
$(D-26)$ replaced with $(D+2N-26)$. This gives the same counting as in the case of the heterotic string
where one fermionic degree of freedom contributes to the central charge as half of a boson.

\section{Acknowledgements}%

We would like to thank Nima Arkani-Hamed, Raphael Flauger, Matthew Kleban, Mehrdad Mirbabayi, Arkady
Tseytlin and Luke Underwood for useful discussions.  This work is supported in part by the NSF under grants
PHY-1068438 and PHY-1316452, and by the NSF CAREER award PHY-1352119.  The work of SS was supported
in part by Fondazione Angelo della Riccia.

\section{Appendix: Scalar Triangle}%
\label{appendix}
After the numerators of triangle diagrams are reduced by the usual procedure we are left with the scalar triangle integral:
\be
I_3=\int\frac{d^dq}{(2\pi)^d}\frac{1}{q^2 (q+p_1)^2 (q+p_1+p_2)^2} \;.
\ee
It was computed in arbitrary dimension in \cite{Bern:1993kr},
\be
I_3=\frac{i}{2}(4 \pi)^{-\frac{d}{2}}\frac{r}{(2-d/2)}\alpha_1\alpha_2\alpha_3\Delta^{\frac{3-d}{2}}\l f(\delta_1)+f(\delta_2)+f(\delta_3)+c \r,
\label{Trig}
\ee
where the following notations are introduced
\be
s_{ij}=(p_i+p_j)^2, \quad p_3=-p_1-p_2, \quad s_{12}=\frac{-1}{\alpha_1 \alpha_3}, \quad s_{23}=\frac{-1}{\alpha_1 \alpha_2}, \quad s_{31}=\frac{-1}{\alpha_2 \alpha_3}\;,
\ee
\be
\gamma_i=\sum_{j=1,3}\alpha_j-2 \alpha_i\;,
\ee
\be
\Delta=\sum_{i>j}\gamma_i \gamma_j\;,
\ee
\be
\delta_i=\gamma_i/\sqrt{\Delta}\;,
\ee

\be
f(\delta)=4^{\frac{d}{2}-1} \delta_{\,2} F_1\l\frac{d}{2}-1,\frac{1}{2};\frac{3}{2};-\delta^2\r\;,
\ee
\be
c=-2\pi\frac{\Gamma(d-3)}{\Gamma \l\frac{d}{2}-1\r^2}\;,
\ee
\be
r=\frac{\Gamma\l 3-\frac{d}{2} \r \Gamma\l \frac{d}{2}-1\r^2}{\Gamma\l d-3 \r}.
\ee
Since (\ref{Trig}) has a pole at $d=2$ we need to know the taylor expansion of the hypergeometric function with respect to its first parameter. In our case it takes the following simple form \cite{Ancarani:2009zz},
\be
{}_2F_1(\epsilon,\frac{1}{2};\frac{3}{2};z)=1+\epsilon \l -2 \frac{\arctanh\sqrt{z}}{\sqrt{z}}-\log(1-z)+2  \r + O(\epsilon^2) \;.
\ee

\bibliographystyle{utphys}
\bibliography{1030}

\end{document}